\documentclass[12pt,draftcls,onecolumn,journal]{IEEEtran}
\usepackage[final]{graphicx}
\usepackage{cite,amsmath,amsthm,amssymb,amsfonts,calrsfs,mathtools}
\usepackage{srcltx}

\newcommand{\comment}[1]{}
\newcommand{\field}[1]{\mathbb{#1}} % requires amsfonts

\newtheorem{theorem}{Theorem}
\newtheorem{corollary}{Corollary}
\newtheorem{lemma}{Lemma}

\makeatletter
\def\ps@headings{%
\def\@oddhead{\mbox{}\scriptsize\rightmark \hfil \thepage}%
\def\@evenhead{\scriptsize\thepage \hfil \leftmark\mbox{}}%
\def\@oddfoot{}%
\def\@evenfoot{}}
\makeatother 
\def\figref#1{Fig.~\ref{#1}}
\def\thref#1{Theorem~\ref{#1}}
\def\corref#1{Corollary~\ref{#1}}
\def\secref#1{Sec.~\ref{#1}}
\def\lmref#1{Lemma~\ref{#1}}

\begin{document}

\title{Wireless Network Code Design and Performance Analysis using Diversity-Multiplexing Tradeoff}

\author{Hakan~Topakkaya and
 Zhengdao~Wang,~\IEEEmembership{Senior Member,~IEEE}
%\thanks{Manuscript received March 23, 2009; }%
\thanks{H. Topakkaya and Z. Wang are with the Department of Electrical and Computer
Engineering, Iowa State University, Ames, IA, 50011, USA (e-mail:
{hakan, zhengdao}@iastate.edu).}}

\maketitle

%\linespread{1}\normalsize
%\renewcommand{\baselinestretch}{2}
\begin{abstract}

Network coding and cooperative communication have received
considerable attention from the research community recently in
order to mitigate the adverse effects of fading in wireless
transmissions and at the same time to achieve high throughput and
better spectral efficiency. In this work, we design and analyze
deterministic and random network coding schemes for a cooperative
communication setup with multiple sources and destinations. We
show that our schemes outperform conventional cooperation in terms
of the diversity-multiplexing tradeoff (DMT). Specifically, it
achieves the full-diversity order at the expense of a slightly
reduced multiplexing rate. We establish the link between the
parity-check matrix for a $(N+M,M,N+1)$ systematic MDS code and
the network coding coefficients in a cooperative communication
system of $N$ source-destination pairs and $M$ relays. We present
two ways to generate the network coding matrix: using the Cauchy
matrices and the Vandermonde matrices, and establish that they
both offer the maximum diversity order.
\end{abstract}
\begin{keywords}
Cooperative communication, network coding, outage probability,
diversity-multiplexing tradeoff.
\end{keywords}

\section{Introduction} \label{intro}

Channel fading is one significant cause of performance degradation
in wireless networks. In order to combat fading, diversity
techniques that operate in time, frequency or space are commonly
employed. The basic idea is to send the signals that carry same
information through different paths, allowing the receiver to
obtain multiple independently faded replicas of the data symbols.
Cooperative diversity tries to exploit spatial diversity using a
collection of distributed antennas belonging to different
terminals, hence creating a virtual array rather than using
physical arrays.

In \cite{acly00} Ahlswede et al. introduced \emph{network coding}
to achieve the max-flow rate for single-source multicast that
could be impossible to achieve by simply routing the data. Since
then, network coding has been recognized as a useful technique in
increasing the throughput of a wired/wireless network. The basic
idea of network coding is that an intermediate node does not
simply route the information but instead combines several input
packets from its neighbors with its own packets and then forwards
it to the next hop. However, since network coding is devised at
the network layer, error-free communication from the physical and
medium-access layer is usually assumed, which is a simplifying
assumption for wireless communications.

Efforts have also been made to apply network coding to the
physical layer, e.g. in \cite{zhll06,zhli09,loyv10}. Towards that
goal, cooperative schemes have been proposed that make use of
network coding in a cooperative communication setup, and studies
have been conducted to determine whether network coding provides
any advantages over existing cooperative communication techniques
\cite{yulb07,pzzy08,hazp07,wagi08,xisk09,woka07,bali08}.

In \cite{pzzy08}, a network-coded cooperation (NCC) was proposed
and its performance was quantified using the
diversity-multiplexing tradeoff analysis which was originally
proposed for multiple antenna systems in \cite{zhts03}. NCC was
shown to outperform conventional cooperation (CC) schemes which
includes space-time coded protocols \cite{lawo03} and selection
relaying \cite{bkrl06}: It requires less bandwidth, and yield
similar or reduced system outage probability while achieving the
same diversity order. However, these results are based on an
optimistic assumption that any destination node should receive the
packets that are not intended for it without any error so that the
intended packet can be recovered from the xor'ed packet sent by
the relay. When this assumption is removed the scheme can no
longer achieve the full diversity order of $M+1$, where $M$ is the
number of cooperating relays, but only a reduced diversity order
of 2.

In this paper, we propose a network coded cooperation schemes for
$N$ source-destination pairs assisted with $M$ relays. The
proposed scheme allows the relays to apply network coding on the
data it has received from its neighbors using the coefficients
from the parity-check matrix of a MDS code. A closed form
expression for the outage probability is derived. We also obtain
the diversity-multiplexing tradeoff performance of the proposed
scheme under two different traffic network models: multicast and
unicast. Specifically, it achieves a maximum diversity order $M+1$
at the expense of a slightly reduced multiplexing rate. We also
propose two different network coding approaches: deterministic and
random.  We show that our scheme outperforms NCC and CC in terms
of probability of outage.

%We use $\field{C}^{1 \times T}$ to denote complex $T$-tuples.

The rest of the paper is organized as follows. Section
\ref{sec.sys} discusses the system model, description of the
proposed scheme. In Section \ref{sec.dmt}, performance analysis is
established using DMT and the main result is presented. Section
\ref{sec.codedesign} discusses the network code design. In Section
\ref{sec.further}, we discuss unicast, random network coding and
selection relaying. In Section \ref{sec.compare} the performance
of the proposed scheme is compared in terms of DMT and average
outage probability with the existing schemes in the literature.
Section \ref{sec.conclusions} contains the conclusions.
\section{System Model} \label{sec.sys}
\subsection{General System Description}

\begin{figure}
\centering
\includegraphics[width=.45\textwidth]{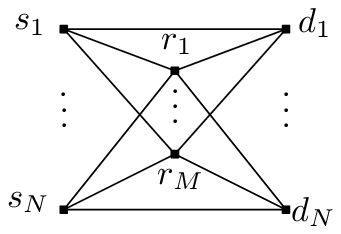}
\caption{System model: $N$ source-destination pairs and $M$
relays} \label{network-cap.pdf}
\end{figure}

The network studied in the paper is composed of $N$
source-destination pairs denoted as $(s_1,d_1), \ldots,
(s_N,d_N)$, and $M$ relays denoted as $r_1, \ldots, r_M$ in a
single-cell where all the nodes can hear the transmissions of each
other as shown in \figref{network-cap.pdf}. We assume that each
packet is composed of $L$ bits: $b_i = [b_{i, 1}, b_{i,2}, \ldots,
b_{i, L}]$. We divide $b_i$ into smaller blocks of equal length
$l$ and represent the $k^{th}$ block $[b_{i, kl+1}, b_{i,kl+2},
\ldots, b_{i,(k+1)l}]$, $k \in \{1, \ldots, K\}$ a finite-field
element $\theta_{i, k} \in \mathbb{F}_q$ where $q=2^l$ and
$K=L/l$. Therefore, each packet is represented as a $K$-tuple
$\Theta_i=[\theta_{i, 1}, \theta_{i,2}, \ldots, \theta_{i,K}] \in
\mathbb{F}_q^{1 \times K}$; see e.g., \cite{lini86},
\cite{drol98}. Dividing each packet into small blocks enables us
to work with a smaller field size which in return significantly
reduces the complexity of the arithmetic operations. This is to be
contrasted to the scheme in \cite{kome03} where the field size is
taken to be $q=2^L$. We will give a lower bound on the field size
in \secref{sec.codedesign}. We consider two different transmission
scenarios. In the first scenario, each source node $s_i$ is trying
to transmit the data packet $\Theta_i$ to all the destinations
$d_i, i=1, \ldots, N$ which is known as the \emph{multicast}
scenario. In the second scenario, each source node $s_i$ is trying
to transmit the data packet $\Theta_i$ to only destination $d_i$
and we will refer to this scenario as the \emph{unicast} scenario.
All the nodes are assumed to be equipped with half-duplex (i.e.
cannot transmit and receive at the same time) single-antennas.
Each data packet $\Theta_i$ is error control coded and modulated,
and transmitted in $T$ time slots.

The channel between any pair of nodes is assumed to be frequency
flat fading with additive white Gaussian noise (AWGN). Let $x_i
\in \field{C}$ denote the transmitted symbols from node $i$ and
$y_j \in \mathbb{C}$ the received symbols at node $j$. The
additive noise $z_i \sim \mathbb{CN}(0,1)$ has independent and
identically distributed (i.i.d.) circularly symmetric entries. Let
$h_{i,j} \in \field{C}$ denote the instantaneous channel
realization. We assume that the channel coefficient $h_{i,j}$
remains constant during the transmission time of a packet. Then,
the channel within one block can be written as
\begin{equation}\label{systemmodel}
  y_j(t)=\sqrt{\rho}h_{i,j}x_i(t)+z_i(t), \quad t=1, 2, \ldots, T.
\end{equation}
where $\rho$ is the average received SNR at the destination. All
the transmissions are made with equal power. In the above
equation, the transmitter could be any of the sources or relays,
the receiver could be any of the relays or destinations, as long
as the transmitter and receiver are different (i.e., not the same
relay). The channel coefficient $h_{i,j}$ between any two nodes is
modeled as i.i.d. with zero-mean, circularly symmetric complex
Gaussian random variables with common variance $1/\beta$.
Therefore, $|h_{i,j}|^2$ is exponentially distributed with
parameter $\beta$ $\forall i,j$ .

A total of $NL$ bits are transmitted by all sources in $(N+M)T$
channel uses, therefore the system rate is $R=NL/[(N+M)T]$ bits
per channel use (BPCU). The transmission rate $R_0$ for one source
or one relay per one packet is fixed, identical, and equal to
$R_0=L/T=R(N+M)/N$ BPCU.

The instantaneous mutual information of the channel model in
\eqref{systemmodel} when i.i.d Gaussian input is used is given by:
\begin{equation}\label{cap}
  I(X_i;Y_j)=\log(1+|h_{i,j}|^2\rho).
\end{equation}
where $X_i$ and $Y_j$ denote the transmitted symbol by node $i$
and received symbol by node $j$. We assume that powerful enough
channel codes can be applied within each packet such that if
$I(X_i;Y_j)>R_0$, the packet can be decoded correctly. In case
errors occur, we assume they can be detected. This can be realized
through cyclic redundancy check (CRC) code or other parity check
codes. When $I(X_i;Y_j) \leq R_0$, we say that the channel
$h_{i,j}$ is in \textit{outage}. Otherwise, we say that the
channel $h_{i,j}$ is \textit{operational}. Define $\tau
=[2^{[R(N+M)]/N}-1]/\rho$. Since $|h_{i,j}|^2$ is exponentially
distributed, the outage probability for the channel in
\eqref{systemmodel} is given by:
 \begin{align}\label{out}
  P_0 =Pr(I(X_i;Y_j)<R_0) =Pr(|h_{i,j}|^2<\tau)
      =1- \exp(-\beta\tau) \cong \beta\tau,
\end{align}
where $a(\tau)\cong b(\tau)$ if $\lim_{\tau\to 0}
[a(\tau)/b(\tau)]=1$.
\subsection{Network Coded Cooperation} \label{main}

\begin{figure*}
\centering
\includegraphics[width=.75\textwidth]{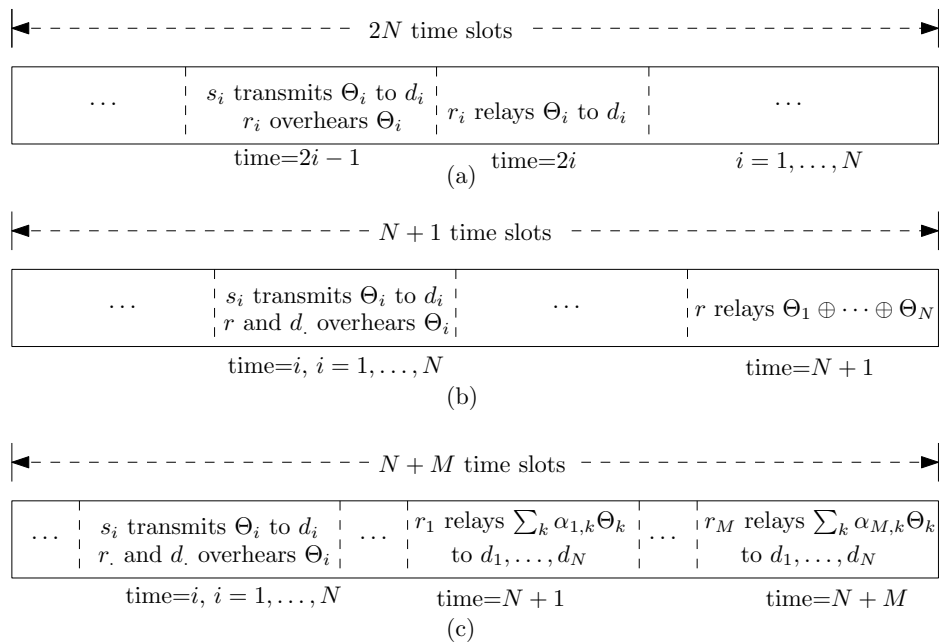}
\caption{time-division allocation for the different schemes
compared: (a) CC (b) NCC (c) DNCC, RNCC} \label{fig.ta}
\end{figure*}

Our transmission scheme consists of two stages; see
\figref{fig.ta}. In the first stage, direct transmissions from the
sources to the destinations take place in $N$ orthogonal time
slots. Thanks to the broadcast nature of the wireless medium, all
the destinations and the relays overhear the transmissions. At the
end of the first stage, each relay tries to decode all $N$
packets. Here one of the two strategies is possible:
\begin{enumerate}
\item{Strategy $\cal{A}$:}
If a relay can successfully decode all the packets, then it
participates in the second stage. Otherwise, it remains silent. In
the second stage, the participating relays perform network coding.
Specifically, relay $i$ will transmit the linear combination $
\sum_{k=1}^{N} \alpha_{ik} \Theta_k.$
\item{Strategy $\cal{B}$:}
If a relay can successfully decode at least one packet, then it
participates in the second stage. Specifically, if relay $i$ was
able to decode the packets correctly from the sources in the set
$S_i$ where $S_i\subseteq \{1,\ldots, N\}$, then it will transmit
the linear combination $ \sum_{k \epsilon S_i} \alpha_{i,k}
\Theta_k$.
\end{enumerate}

Unless otherwise specified, we study the first case when the
Strategy $\cal{A}$ is used until \secref{sec.further}. Strategy
$\cal{B}$ will be discussed in \secref{sec.further}.

\subsection{Deterministic and Random Network Coding}
We will consider two network coding schemes for the user
cooperation: random coding and deterministic coding. In the random
coding approach, which we will refer to as Random Network Coded
Cooperation (RNCC), relay $r_i$ draws $\alpha_{ij}$ randomly from
the finite field $\mathbb{F}_q$. After the random coefficients are
drawn, a new packet is created by making a linear combination of
the source data packets using the $\alpha_{ij}$'s. In the
deterministic approach which will be referred to as Deterministic
Network Coded Cooperation (DNCC), the coefficients $\alpha_{ij}$'s
are predetermined and they are designed in a way to maximize the
probability that the received linear combinations are actually
decodable at the destination. We will discuss the problem of how
to choose these predetermined coefficients in detail in
\secref{sec.codedesign}.

In order to express the overall transmitted signal, we define the
following matrix:
\begin{equation} \label{eq.matrixA}
 A  := \begin{bmatrix}
     1 & \ldots        & 0  & \alpha_{1,1}  & \ldots &  \alpha_{M,1}\\
     \vdots & \ddots        & \vdots  & \vdots             &     .  &  \vdots \\
     0 & \ldots        & 1  & \alpha_{1,N}  & \ldots &  \alpha_{M,N}
  \end{bmatrix}^T
\end{equation}
where $(\cdot)^T$ denotes transpose. Also define the $N \times K$
finite field vector corresponding to the original source packets
as $\Theta=[\Theta_1^T, \Theta_2^T, \ldots, \Theta_N^T]^T$.
%\begin{equation}
%\Theta=[\Theta_1^T, \Theta_2^T, \ldots, \Theta_N^T]^T
%%\Theta=[\Theta_1, \Theta_2, \ldots, \Theta_N]^T
%\end{equation}
Using matrices $A$ and $\Theta$, we can express the potential
transmitted signals by all the $N$ sources and $M$ relays, in that
order, as $\Pi=A\Theta$ where $\Pi \in \mathbb{F}_q^{(N+M) \times
K}$. Note that $\Pi$ represents the \textit{potential} transmitted
signals, since due to severe fading some of the channels might be
in outage and therefore only a subset of packets can be
successfully decoded by some relays. Under Strategy $\cal{A}$,
such relays will not participate in the second stage and the rows
of $A$ corresponding to these relays can be considered to be
deleted. Under Strategy $\cal{B}$, however, only the coefficients
in $A$ that correspond to the unsuccessful packets would be zero,
as opposed to a whole row being deleted. Note that, from the
destination $d_i$'s perspective, some of the channels might also
be in outage. We denote the corresponding submatrix of $A$ for
destination $d_i$ by $A_i$ which satisfies $\Pi_i=A_i\Theta$ where
$\Pi_i$ denotes all correctly decoded packets at destination
$d_i$.
\section{Performance Analysis} \label{sec.perf}
\subsection{Diversity-Multiplexing Tradeoff} \label{sec.dmt} As
mentioned in the introduction, we will investigate the performance
of the proposed scheme via diversity-multiplexing tradeoff (DMT).
DMT is accepted as a useful performance analysis tool in
cooperative systems \cite{pzzy08,bkrl06}. For completeness, we
give the formal definitions as in \cite{zhts03}. Let $P_e^i(\rho)$
denote packet error probability of user $i$ at SNR $\rho$. Define
$P_e=\min_i P_e^i$, $i=1, \ldots, N$, then a scheme is said to
achieve spatial multiplexing gain $r$ and diversity gain $d$ if
the data rate is $\lim_{\rho \rightarrow
\infty}R(\rho)/\log(\rho)=r$,
%\begin{equation}\label{rn}
%  \lim_{\rho \rightarrow \infty} \frac{R(\rho)}{\log(\rho)} = r ,
%\end{equation}
and the minimum error probability satisfies $\lim_{\rho
\rightarrow \infty}{\log(P_e(\rho))/\log(\rho)}=-d$.
%\begin{equation}\label{d}
%  \lim_{\rho \rightarrow \infty} \frac{\log(P_e(\rho))}{\log(\rho)} = -d .
%\end{equation}
%In \cite{zhts03}, for long enough block length $T$ it was shown
%that the probability of error is dominated by the outage
%probability in terms of the diversity order. Therefore, we will
%only consider outage probabilities in our analysis.

\subsection{Main Result}
Next we define a new parameter which plays a key role in the
derivation of the outage probability and hence the achieved
diversity order. For any integer $i \in [1, \min(m,n)]$, we define
the $\Gamma$-rank, $\Gamma_i(C)$, of a $m \times n$ matrix $C$ as
an integer $\gamma$ such that $1)$ any collection of $\gamma$ rows
of $C$ is at least rank $i$, and $2)$ there exists a collection of
$\gamma -1$ rows of $C$ that has rank $i-1$. Next, we derive the
DMT of the system as a function of $\Gamma_N(A)$.
\begin{theorem} \label{th.beta}
The diversity-multiplexing tradeoff of DNCC with $N$
source-destination pairs and $M$ intermediate relay nodes which
choose their linear combination coefficients from the matrix $A$
for multicast using Strategy $\cal{A}$ is given by:
\begin{equation}\label{dmtbeta}
 d\left(r\right) = \left(N+M-(\Gamma_N(A) -1)\right)\left[1-\frac{N+M}{N}r\right],
\quad  r \in \left(0,\frac{N}{N+M}\right).
\end{equation}
\end{theorem}
\begin{proof}\label{app.th.beta}
\subsubsection{Multicast}
In the multicast problem, the necessary and sufficient condition
for destination $d_i$ to recover $\Theta_i$ is $rank(A_i)=N$. To
analyze the outage probability, we define the following events:
$E_{i} \triangleq \{ rank(A_i) < N \}$, and $E_{i}^{up} \triangleq
\{ A_i$ has at most $\Gamma_N(A)-1$ rows\}. Notice that, $E_i
\subset E_i^{up}$ by the first condition in the definition of
$\Gamma$-rank. By the second condition in the definition of
$\Gamma$-rank, there exist a collection of rows of $A$ that are
rank $N-1$. Let $\tilde{A}_i$ denote a $(\Gamma_N(A)-1) \times N$
submatrix of $A$ that consists of such rows. Let $F_m$ denote the
event that $m$ relays fail to receive all the $\Theta_i$'s
correctly. Define $E_i^{low} \triangleq \{ F_0 \cap
\{A_i=\tilde{A}_i\}\}$. It follows that $E_i^{low} \subset E_i$.
Notice that the probability that any relay can successfully decode
all $N$ packets in the first stage is $P(S) = \prod_{i=1}^{N}
{Pr(I_{s_ir}(X;Y)>R_0) }
  = \prod_{i=1}^N \exp(-\beta\tau)  = \exp(-N\beta\tau)$. As a result,
\begin{equation}\label{F_m}
 P(F_m) = \binom Mm P(S)^{M-m} (1-P(S))^m.
\end{equation}

Having $N$ direct transmission from the sources and $M-m$
transmissions from the relays, each destination can potentially
receive and decode $N+M-m$ packets. Let $E(N+M-m,l)$ denote the
event that $l$ out of $N+M-m$ channels were operational:
\begin{align}\label{E(k,l)}
 P(E(N+M-m,l)) = \binom {N+M-m}l  P_0^{N+M-m-l}(1-P_0)^{l}
 \end{align}
where $P_0$ is given by \eqref{out}. Since $E_i^{low} \subset E_i
\subset E_i^{up}$, using \eqref{F_m} and \eqref{E(k,l)} we have:
\begin{align}\label{error1iupper}
  P(E_{i}) \leq P(E_{i}^{up})=\sum_{m=0}^{M} {P(F_m)} \cdot
  {\sum_{l=0}^{\Gamma_N(A)-1} P(E(N+M-m,l))}
\shortintertext{and}
 P(E_{i}) \geq P(E_{i}^{low})=P(F_0) P_0^{N+M-(\Gamma_N(A)-1)}(1-P_0)^{\Gamma_N(A)-1}
 \end{align}
In \eqref{error1iupper}, the first summation stands for the
probability of the event that $m$ of the relays fail to receive
all $\Theta_i$'s correctly, leaving us with only $M-m$ relays
which will participate in the second stage. In total $N+M-m$
transmissions will be made. The destination $d_i$ may not be able
to recover all $\Theta_i$'s, if only $\Gamma_N(A)-1$ or less
number of transmissions are successful.

Notice that, as $\rho \rightarrow \infty$, $\tau \rightarrow 0$.
We need to find the following limit:
  \begin{equation}\label{limit}
 \lim_{\tau \rightarrow
 0}{\frac{P(E_{i})}{\tau^{N+M-(\Gamma_N(A)-1)}}}.
  \end{equation}

We consider the individual terms in the summations one-by-one and
find the term with the smallest order of $\tau$. Observe that
$\lim_{\tau \rightarrow 0}{(1-P(S))}=N\beta$ and $P(F_m) \cong
K_m\tau^m $ where $K_m$ is:
\begin{align}\label{limitFm}
  \lim_{\tau \rightarrow 0}{\frac{P(F_m)}{\tau^{m}}}
              = \binom Mm (N\beta)^m.
  \end{align}

Similarly, $P(E(k,l)) \cong K_{k,l}\tau^{k-l} $ where
$K_{k,l}=\binom kl \beta^{k-l}$. The smallest order $\tau$ term
happens when $l$ is equal to $\Gamma_N(A)-1$. Hence, we have:
  \begin{align}\label{eq.K}
   \lim_{\tau \rightarrow 0}
  \frac{P(F_m)}{\tau^{m}}\cdot
   {\frac{P(E(N+M-m,l))}{\tau^{N+M-m-(\Gamma_N(A)-1)}}}
  = K_m K_{N+M-m,\Gamma_N(A)-1} \triangleq K_{up}, \\
\shortintertext{and}
 P(E_{i}^{up})  \cong K_{up} \tau^{N+M-(\Gamma_N(A)-1)}  =  K_{up}
              \left({\frac{2^{\frac{N+M}{N}R}-1}{\rho}}\right)^{N+M-(\Gamma_N(A)-1)}.
   \end{align}
Similarly, we can show that $P(E_{i}^{low}) \cong K_{low}
\tau^{N+M-(\Gamma_N(A)-1)}$ where
$K_{low}=\beta^{N+M-(\Gamma_N(A)-1)}$. Now, choosing the fixed
rate to be $R=r\log \rho$ and substituting into \eqref{eq.K}, we
obtain:
   \begin{align}\label{final}
 P(E_{i})  \cong K \rho^{(\frac{N+M}{N}r-1)(N+M-(\Gamma_N(A)-1))}
   \end{align}
where $K_{low} \leq K \leq K_{up}$, which is the desired result.
\subsubsection{Unicast}
In unicast we have a different problem: given the received packets
$Y_i$ at destination $d_i$, we would like to recover only
$\Theta_i$ from the set of linear equations $Y_i=A_i\Theta$. The
error can only happen when the direct link is in outage. Notice
that this implies that $A_i$ does not contain $e_i$ (the $i$'th
row of the $N\times N$ identity matrix $I_{N \times N}$). In this
case, a necessary and sufficient condition for $\Theta_i$ to be
recoverable is that $e_i \in span(A_i)$, where $span(A_i)$ is the
row-space of $A_i$. Here, we make another rank definition that
will be useful for the proof of the unicast scenario. We define
the $\Lambda_i$-rank, $\Lambda_i(C)$ of a $m \times n$ matrix $C$
as an integer $\lambda$ such that $1)$ any collection of $\lambda$
rows of $C$ spans a space that contains $e_i$ but $2)$ there
exists a collection of $\lambda -1$ rows of $C$ which does not
span a space that contains $e_i$. Next, we derive the DMT of the
system as a function of $\Lambda_i(A)$.
\begin{lemma} DMT of DNCC for unicast for the $i^{th}$ destination is
\begin{equation}\label{dmtlambda}
 d_i\left(r\right) = \left(N+M-(\Lambda_i(A) -1)\right)\left[1-\frac{N+M}{N}r\right],
\quad  r \in \left(0,\frac{N}{N+M}\right).
\end{equation}
\end{lemma}
\begin{proof}
Here we define the following relevant events. $\bar{E_{i}}
\triangleq \{ e_i \not \in span(A_i)\}$, $\bar{E_{i}}^{up}
\triangleq \{ A_i$ has at most $\Lambda_i(A)-1$ rows\}. Notice
that, $\bar{E_i} \subset \bar{E_i}^{up}$ by the first condition in
the definition of $\Lambda_i$-rank. By the second condition in the
definition of $\Lambda_i$-rank, there exist a collection of
$\Lambda_i(A)-1$ rows of $A_i$ that does not span $e_i$. Let
$\bar{A}_i$ denote a $(\Lambda_i(A)-1) \times N$ submatrix of
$A_i$ that consists of such rows. Keeping the definition of $F_m$
define $\bar{E_i}^{low} \triangleq \{ F_0 \cap
\{A_i=\bar{A}_i\}\}$. It follows that $\bar{E_i}^{low} \subset
\bar{E_i}$. Since $\bar{E_i^{low}} \subset \bar{E_i} \subset
\bar{E_i}^{up}$, using \eqref{F_m} and \eqref{E(k,l)} we have:
\begin{align}\label{upandlow}
  P(\bar{E_{i}}) \leq P(\bar{E_{i}}^{up})=P_0\sum_{m=0}^{M} {P(F_m)} \cdot
  {\sum_{l=0}^{\Lambda_i(A)-1} P(E(N-1+M-m,l))}
\shortintertext{and}
 P(\bar{E_{i}}) \geq P(\bar{E_{i}}^{low})=P(F_0) P_0^{N+M-(\Lambda_i(A)-1)}(1-P_0)^{\Lambda_i(A)-1}
 \end{align}
where the first $P_0$ in \eqref{upandlow} accounts for the outage
of the direct link between $s_i$ and $d_i$. The limits in the
second summation in \eqref{upandlow} is due to the fact that the
destination $d_i$ may not be able to recover all $\Theta_i$'s, if
only $\Lambda_i(A)-1$ or less number of transmissions are
successful. The rest of the proof can be completed by showing that
the diversity orders of both the upper and the lower bound are
equal to \eqref{dmtlambda} as in the proof of the Multicast
scenario.
\end{proof}
Notice that from the definition of $\Gamma$-rank, we have
$\Gamma_N(A)=\max_i{\Lambda_i(A)}$. Since the error probability
$P_e$ is defined to be the minimum of individual error
probabilities, we have $d(r)=\min_i d_i(r)$. After substituting
$d_i(r)$ in \eqref{dmtlambda}, we obtain the desired result in
\eqref{dmtbeta}.
\end{proof}
\begin{corollary}\label{cor.A} The maximum diversity
is achieved if and only if $\Gamma_N(A)=N$.
\end{corollary}
\begin{proof}\label{app.cor}
The result follows immediately from \eqref{dmtbeta}.
\end{proof}

\section{Design of the Linear Network Coding Matrix} \label{sec.codedesign}

In this section, we try to design a network coding matrix $A$ that
can yield the maximum diversity order. Notice that, by definition
we have $\Gamma_N(A) \geq N$. Therefore, it is clear from
\corref{cor.A} that we need to pick an $A$ that satisfies
$\Gamma_N(A)=N$. Before going into the discussion on the design of
the matrix $A$, we would like to give another important rank
definition that will be used in the design of $A$.

The \emph{row Kruskal-rank} \cite{sigb00,wagi01i} of $A$, denoted
by $\kappa(A)$, is the number $r$ such that every set of $r$ rows
of $A$ is linearly independent, but there exist one set of $r + 1$
rows that are linearly dependent.
\begin{lemma}\label{ranksequal}
$\kappa(A)=N \Leftrightarrow \Gamma_N(A)=N$.
\end{lemma}
\begin{proof}
We prove $\Gamma_N(A)=N \Rightarrow \kappa(A)=N$; the other case
is straightforward. When $\Gamma_N(A)=N$, from the definition of
$\Gamma$-rank any collection of $N$ rows of $A$ is at least rank
$N$. But since $rank(A) \leq N$, we have the first condition for
the Kruskal-rank. Also since $rank(A) \leq N$, any $N+1$ rows will
be linearly dependent.
\end{proof}
The minimum Hamming distance $d_{min}$ between any two codewords
for a $(n,k)$ error-correcting code is upper bounded by the
Singleton bound as $d_{min} \leq n-k+1$. The codes that achieve
this bound are called maximum distance separable (MDS) codes
\cite{masl77}. The following result relates the column
Kruskal-rank of the parity-check matrix $H$ of a linear block code
to its minimum distance $d_{min}$: $d_{min}=\kappa(H) + 1$. This
follows from the following theorem in \cite{masl77} by realizing
that $d_{min}=n-k+1$ for an $[n,k,d]$ MDS code $\cal{C}$.
\begin{theorem}\label{th.Kruskaldmin}
(\cite[p.~318]{masl77}) $\cal{C}$ is MDS if and only if (iff)
every $n-k$ columns of $H$ are linearly independent.
\end{theorem}

The transpose $H^T$ of the parity check matrix of a systematic
$(N+M,M,N+1)$ MDS code can be used as an encoding matrix $A$ for
our DNCC scheme to minimize the total number of packets necessary
at the destinations for decoding the source packets. If such an
$A$ is used, then each destination needs and only needs $N$
packets (from the sources and relays) for correct decoding. Note
that depending on the sizes $N$ and $M$, finding a $(N+M,M,N+1)$
MDS code may or may not be possible in a given finite field
$\mathbb{F}_q$ \cite{masl77,sing64}.
\subsection{Network Code Designs from RS Codes}
Reed-Solomon (RS) Codes are MDS codes. There are two ways of
constructing an RS Code: either using the Vandermonde matrices
\cite{reso60} or using the Cauchy matrices
\cite[Sec.~4.3]{wibh99}. Because of the special structure of $A$
in \eqref{eq.matrixA}, we will be working on systematic RS codes.
\subsubsection{Construction based on Cauchy
Matrices} The systematic generator matrix for the RS$(n,k)$ code
has the form $G=[I|C]$ where $I$ is the identity matrix of order
$k$ and $C$ is a $k \times (n-k)$ matrix \cite{wibh99} and $G$
satisfies $\kappa(G^T)=n-k$. $C$ is known as Cauchy matrix and is
given by:
\begin{equation}
C_{i,j}=\frac{u_iv_j}{x_i+y_j}, \quad 0 \geq i \geq k-1, \quad 0
\geq j \geq n-k-1.
\end{equation} where $u_i, v_j, x_i$ and $y_j$ are elements of
$GF(2^m)$ and are defined as:
\begin{align}
x_i&=\beta^{n-1-i}, \quad 0 \geq i \geq k-1,  \\
y_j&=\beta^{n-1-k-j}, \quad 0 \geq j \geq n-k-1,  \\
u_i&=\frac{1}{\prod_{0 \geq i \geq k-1, \; l \neq
i}{(\beta^{n-1-i}-\beta^{n-1-l})}}, \quad 0 \geq i \geq k-1, \\
%\end{align}
%\begin{align}
v_j&=\prod_{0 \geq l \geq k-1}(\beta^{n-1-k-j}-\beta^{n-1-l}),
\quad 0 \geq j \geq n-k-1.
\end{align}
where $\beta$ is the primitive element for $\mathbb{F}_q$.
Therefore, choosing $n=N+M$ and $k=M$, we construct the network
code $A=[I|\alpha]$ by choosing $\alpha_{i,j}=C_{i,j}$ which gives
$\kappa(A)=N$.
\subsubsection{Construction based on Vandermonde
Matrices} The Vandermonde matrices are defined from a vector of
$m$ distinct generating elements $\{ t_1, \ldots, t_m\}$ of
$\mathbb{F}_q$ as:
\begin{equation} \label{eq.Vandermonde}
 V_{m \times n}  := \begin{bmatrix}
     1 & t_1 & t_1^2 & \ldots &  t_1^{n-1}\\
     . & .   & .     & \ldots & .\\
     . & .   & .     & \ldots & . \\
     1 & t_m & t_m^2 & \ldots & t_m^{n-1}
 \end{bmatrix}.
\end{equation}
The determinant for the square Vandermonde matrix is given by
$\det(V_{n \times n})=\prod_{1 \leq i < j \leq n}(t_j-t_i))$ and
$V_n$ is nonsingular iff all the $t_i$'s are distinct. To
construct $A$ for given $N$ and $M$, we do the following:
\begin{enumerate}
\item Choose a suitable
$\mathbb{F}_q$ with $q=2^l \geq N+M$.
\item Choose $N+M$ distinct elements $t_1, t_2, \ldots, t_{N+M}$ of
$\mathbb{F}_q$.
\item Construct the Vandermonde matrix $V_{N \times N}$ from $t_1, \ldots, t_{N}$ and
$V_{M \times N}$ from $t_{N+1}, \ldots, t_{N+M}$.
\item Then $\alpha_{i,j}=(V_{M \times N}V_{N \times
N}^{-1})_{i,j}$ and $A=[I|\alpha^T]^T$.
\end{enumerate}

Note that the generating elements that are needed in the
construction of RS codes from Vandermonde matrices requires an
extra property that they should be the consecutive powers of the
primitive element $\beta \in \mathbb{F}_q$, i.e. $\beta, \beta^2,
\ldots, \beta^{2t}$ for a $t$-error correcting RS code. We do not
need or impose this requirement.
\begin{lemma}\label{lem.Vandermonde} Let $T \in
\mathbb{F}_q^{n \times n}$ be an invertible matrix. Then, for any
$H \in \mathbb{F}_q^{m \times n}$ is the same as that of $HT$:
$\Gamma_i(H)=\Gamma_i(HT)$ and $\kappa(H)=\kappa(HT)$, $\forall \;
1 \leq i \leq \min(m,n)$.
\end{lemma}
\begin{proof}
Pick $i \in (1, \min(m,n))$ arbitrary rows from $H$ and denote the
resulting $n \times n$ matrix by $H'$. We need to show that if the
rows of $H'$ are linearly dependent or linearly independent then
so are the rows of $H'T$. But notice that, since $T$ is full-rank
we have $xH'=0 \Leftrightarrow xH'T=0$ where $x \in
\mathbb{F}_q^N$ and $\mathbb{F}_q^N$ is the $N$-tuples in
$\mathbb{F}_q$.
\end{proof}
Picking $H=[V_{N \times N}|V_{M \times N}]$ and $T=V_{N \times
N}^{-1}$ and using \lmref{lem.Vandermonde} we have
$\Gamma_N(G)=\kappa(G)=N$. Note that since we need $N+M$ distinct
elements of the finite field $\mathbb{F}_q$, it is enough to have
$q \geq N+M$. Next, we give an example for the case when $N=2$ and
$M=2$.

%\begin{example}
\textbf{Example:} Consider a $[n,k,d]=[4,2,3]$ MDS code
$\mathcal{C}$ over
$\mathbb{F}_4=\{0,1,\alpha,\beta=\alpha^2=\alpha+1 \}$ with symbol
representations as $\{0=(0,0),1=(0,1),\alpha=(1,0),\beta=(1,1)
\}$. Constructing the Vandermonde matrices from the set
$\{0,1,\alpha,\beta\}$ and multiplying with the inverse, we have
\begin{equation}
V=\left(%
\begin{array}{cccc}
  1 & 1 & 1      & 1 \\
  0 & 1 & \alpha & \beta \\
\end{array}
\right)^T \quad A=\left(%
\begin{array}{cccc}
  1 & 0 & \beta  & \alpha \\
  0 & 1 & \alpha & \beta \\
\end{array}\right)^T.
\end{equation}
%$\{0,1,\alpha,\beta\}$ $V=\left(%
%\begin{array}{cccc}
%  1 & 1 & 1      & 1 \\
%  0 & 1 & \alpha & \beta \\
%\end{array}
%\right)^T$ and multiplying with the inverse, we have $A=\left(%
%\begin{array}{cccc}
%  1 & 0 & \beta  & \alpha \\
%  0 & 1 & \alpha & \beta \\
%\end{array}\right)^T$.
Let $K=1$ and $b_i = [b_{i, 1}, b_{i,2}], \Theta_i=[\theta_{i,1}],
i=\{1,2\}$. If the above encoder matrix is used, relays will
transmit the linear combinations
$\beta\theta_{1,1}+\alpha\theta_{2,1}, \alpha \theta_{1,1} + \beta
\theta_{2,1}$ respectively. For example for the first relay, this
operation will be performed using regular addition in
$\mathbb{F}_2$ as
$(b_{1,2}+b_{2,1}+b_{2,2},b_{1,1}+b_{1,2}+b_{2,1})$ and for the
second relay $(b_{1,1}+b_{1,2}+b_{2,2},b_{1,1}+b_{2,1}+b_{2,2})$
which can be put in the matrix form as:
\begin{equation} \label{eq.matrixformNC}
\left(%
\begin{array}{cccc}
  0 & 1 & 1 & 1 \\
  1 & 1 & 1 & 0 \\
  1 & 1 & 0 & 1 \\
  1 & 0 & 1 & 1 \\
\end{array}
\right) \left(%
\begin{array}{c}
  b_{1, 1} \\
  b_{1, 2} \\
  b_{2, 1} \\
  b_{2, 2} \\
\end{array}%
\right). \end{equation}.

Notice that since $\kappa(A)=2$ and the destination will be able
to recover $\Theta_1, \Theta_2$ when at least two of the
transmissions are successful. It is important to emphasize that
unlike MDS code construction that we gave earlier, here we do not
have any restrictions on the code size for any given $N$ and $M$,
we can find a large enough finite field $\mathbb{F}_q$ satisfying
$q=2^l, l \geq N+M$.
\section{Discussions and Further Improvements} \label{sec.further}
In the previous sections, we have established the DMT of DNCC for
a given network coding matrix $A$. Later, we designed the network
coding matrix $A$ to have the property that
$\Gamma_N(A)=\kappa(A)=N$. Here in this section, we look at the
performance of DNCC under the Strategy $\cal{B}$ and the case when
only some of the relays (which are selected according to their
channel qualities) are allowed to transmit. We also investigate
the performance of the case when the linear combination
coefficients are chosen randomly.
\subsection{Strategy $\cal{B}$} Decode-and-forward schemes suffer
from performance loss when the source-relay channel is in outage.
If a multi-source scenario is considered the performance loss
becomes even more severe. Therefore, the assumption that the relay
has to decode all the packets in order to be able to cooperate may
be too restrictive for such schemes. We could relax this
assumption and assume that the relays will participate cooperation
even though they have not been able to decode all the packets.

Denote the outage event under Strategy $\cal{B}$ by
$E_{i}^{\cal{B}}$. Notice that under Strategy $\cal{B}$, not only
the $M-m$ relays as in \eqref{F_m} but also the rest of the $m$
relays contributes to the decoding at the destinations. Clearly
the probability of not being able to solve the linear system of
equations will decrease and hence the performance will get better,
i.e. $P(E_{i}^{\cal{B}}) \leq P(E_{i})$. Taking $\Gamma_N(A)=N$,
the probability of $E_i^{low}$ becomes
$P(E_i^{low})=P(F_0)P_0^{M+1}(1-P_0)^{N-1}$. We have $P(E_i^{low})
\leq P(E_{i}^{\cal{B}}) \leq P(E_{i})$. But using a similar
analysis as in the proof of \thref{th.beta}, it can be shown that
even though Strategy $\cal{B}$ offers lower packet error rate, the
DMT is the same as that of Strategy $\cal{A}$. That is, even
though Strategy $\cal{B}$ improves the packet error rate
performance, the DMT remains unchanged.
\subsection{RNCC:} In RNCC the linear combination coefficients
$\alpha_{i,j}$'s are chosen randomly from a finite field
$\mathbb{F}_q$. Similar to the deterministic case, destination
$d_i$ cannot recover $\Theta_i$ when the submatrix $A_i$ is rank
deficient, i.e. $E_i=rank(A_i)<N$. However, unlike the
deterministic case there are two possible reasons to have a rank
deficient $A_i$ in the random case: one is due to fading and the
other is due to the choice of the random coefficients
$\alpha_{ij}$'s. The former happens when at most $N-1$ channels
are operational resulting in an $A_i$ matrix that has at most
$N-1$ rows. Notice that no matter what $\alpha_{i,j}$'s are chosen
$A_i$ will be rank deficient and hence the linear system of
equations cannot be solved. Therefore, we define this event to be
deterministic error event: $E_{i}^{det}=\{A_i$ has at most $N-1$
rows\} in the proof of \thref{th.beta} and probability of this
event is given by $P(E_{i}^{det})=\sum_{m=0}^{M} {P(F_m)} \cdot
  {\sum_{l=0}^{N-1} P(E(N+M-m,l))}$.
Notice that $E_{i}^{det} \subset E_i$. On the other hand, due to
the random choice, relays may choose linearly dependent
coefficients which will result in an $A_i$ matrix such that
$rank(A_i)<N$. Denote this event by $E_i^{ran}$. But by the
\corref{cor.A}, this event will result in the outage events that
have diversity order less than $M+1$. Therefore, the key idea of
this proof is to isolate such events, and show that the
probability of such events can be bounded by the field size.
\begin{align}\label{piRNCC}
P(E_i) & =
P(E_i|E_{i}^{det})P(E_{i}^{det})+P(E_i|E_i^{ran})P(E_i^{ran}) \nonumber \\
& = P(E_{i}^{det}) + P(E_i|E_i^{ran})P(E_i^{ran})  \leq
P(E_{i}^{det}) + P(E_i|E_i^{ran})
\end{align}

Next, we present the lemma that upper bounds the term
$P(E_i|E_{i}^{ran})$.
\begin{lemma}\label{lemmapenalty} The probability that any $N \times N$
square submatrix $A_{i}^\prime$ of $A_{i}$ is rank deficient is
upper bounded by,
\begin{align}\label{det}
P(E_i|E_i^{ran}) \leq \frac{N}{q}
\end{align}
%where $q$ is the field size.
\end{lemma}
\begin{proof}
The proof consists of an application of the Schwartz-Zippel Lemma
of a carefully chosen error event as in the proof of Theorem 2 in
\cite{hmkk06}. We skip the proof due to space limitations.
%The worst case scenario happens when all the channels between the
%sources and the destination are in outage and $A_{i}^\prime =
%A_{i}$. This is the worst case scenario since if $k$ sources are
%not in outage, then we only need to consider the determinant of
%$(N-k) \times (N-k)$ submatrix of $A_i$ whose probability of being
%rank deficient is less than that of $N \times N$ matrix. And since
%$A_i$ has at least $N$ rows when we condition on the event
%$E_{i}^c$, it is clear that the probability of having a rank
%deficient $A_i$ matrix in the case when $A_{i}^\prime \neq A_{i}$
%is smaller than the case when $A_{i}^\prime = A_{i}$. Then
%considering the case when $A_{i}^\prime = A_{i}$, since the
%determinant of $A_{i}$ is a polynomial in terms of the
%indeterminant $\alpha_{ij}$'s with degree $N$, using
%Schwartz-Zippel Lemma we have the desired result. The bound can be
%further improved as in [Lemma-4,\cite{hmkk06}] to be
%$1-\left(1-\frac{1}{q}\right)^N$.
\end{proof}

Now, using \lmref{lemmapenalty} we see that if the field size is
large enough, the error event will be dominated by the event
$E_{i}$ with high probability:
\begin{align}
\lim_{q\rightarrow \infty} P(E_i) \leq P(E_i^{det}) +
\lim_{q\rightarrow \infty} \frac{N}{q} = P(E_i^{det}).
\end{align}

Note that, although the limit is taken asymptotically with
$q\rightarrow \infty$, it is enough to have $q \cong \rho^{M+1}$.
The rest of the proof is the same with the above proof for DNCC.
Below we summarize all the above proved results with the following
theorem:
\begin{theorem} \label{th.maindmt}
DNCC with $M$ intermediate relay nodes which choose their linear
combination coefficients from the rows of $A$ that satisfies
$\Gamma_N(A)=N$ and $N$ source nodes achieves the DMT in both the
multicast and unicast scenario and under both strategies $\cal{A}$
and $\cal{B}$:
\begin{equation}\label{dmtdncc}
 d\left(r\right) = \left(M+1\right)\left[1-\frac{\left(N+M\right)}{N}r\right],
  r \in \left(0,\frac{N}{N+M}\right).
\end{equation}
RNCC achieves the same DMT as in \eqref{dmtdncc} with probability
at least $1-\frac{N}{q}$, where q is the field size.
\end{theorem}

\subsection{Selection Relaying:} We can also consider the case
where not all of the $M$ relays transmit, but only $K$ selected
relays transmit. The same selection rule based on the
instantaneous wireless channel conditions can be adapted as in
\cite{pzzy08}. Define
%$h_i=\min\{|h_{{s_{1i}}{r_i}}|^2,|h_{{r_i}{d_{1i}}}|^2,\ldots,
%|h_{{s_N}{r_i}}|^2,|h_{{r_i}{d_N}}|^2\}$
\begin{equation}\label{selectionrule}
h_i \triangleq
\min\{|h_{{s_{1i}}{r_i}}|^2,|h_{{r_i}{d_{1i}}}|^2,\ldots,
|h_{{s_N}{r_i}}|^2,|h_{{r_i}{d_N}}|^2\}
\end{equation}
where $h_{i,j}$ is the channel coefficient between node $i$ and
node $j$. Then, select the $K$ relays that maximizes $h_i$, namely
first choose $r$ with the rule: $r=\arg \max_{r_i}{h_i}$
%\begin{equation}\label{best}
%r=\arg \max_{r_i}{h_i}
%\end{equation}
and continue the same process of choosing the maximum in the
beginning of each relay transmission. Note that this selection
mechanism can be implemented using a distributed protocol at the
network layer as in \cite{bkrl06}: relays choose a timer that is
inversely proportional to the quality of their channels. Relays
assess the quality of their channels from the RTS-CTS packets that
were transmitted by the source and destination nodes
respectively\footnote{Forward and backward channels between the
relays and the destinations are assumed to be the same due to
reciprocity theorem \cite{rapp96}.}. No CSI is required at the
physical layer. Next, we give the diversity-multiplexing
performance of this scheme.
\begin{theorem} \label{extension}
DNCC scheme with the selection of the best $K$ relay nodes out of
$M$ and $N$ source nodes, $\Gamma_N(A)=N$ in the multicast
scenario achieves the DMT:
\begin{equation}
 d\left(r\right) = \left(K+1\right)\left(1-\frac{\left(N+K\right)}{N} r\right), \quad r \in \left(0,\frac{N}{N+K}\right)
 \end{equation}
if $K<N-1$, and otherwise achieves the DMT:
\begin{equation}
 d\left(r\right) = \left(N+M\left(K-(N-1)\right)\right)\left(1-\frac{N+K}{N} r\right), \quad r \in \left(0,\frac{N}{N+K}\right)
 \end{equation}
 \end{theorem}
\begin{proof}
Let $r= \arg \max{h_i}$ where $h_i$ is as in
\eqref{selectionrule}. The cdf for $|h_{jr}|^2$ (or $|h_{rj}|^2$)
where $j$ can be a source (or a destination) node was derived in
\cite{pzzy08} as:
\begin{align}\label{sel.cdf}
& F\left(\tau\right)=\int_0^{\tau}\sum_{m=1}^M
\left\{\beta_{kr_m}e^{-\beta_m
\phi} \prod_{j \neq m}^M\left(1-e^{-\beta_{j}}\right)\right\}d\phi +   \nonumber \\
& \int_0^{\tau} \sum_{m=1}^M \int_o^{\phi}
\left(\beta_m-\beta_{mk}\right)e^{-\left(\beta_m-\beta_{mk}\right)\theta}\prod_{j
\neq m}^M \left(1-e^{-\beta_{j}\theta}\right)d\theta
\beta_{mk}e^{-\beta_{mk}\phi}d\phi
 \end{align}
where $\beta_{m,k}$'s are the parameters of the exponential random
variables associated with the corresponding channels between node
$m$ and node $k$, and
$\beta_m=\sum_{k=1}^{N}\left[\beta_{m,k}+\beta_{k,m}\right],
m=\{1, \ldots, M\}$. Taking $\beta_{m,k}=\beta$ and using
exponential expansion a high-SNR approximation for \eqref{sel.cdf}
can be shown to be equal to: $F(\tau) \cong (2N \beta)^{M-1} \beta
\tau^M$. Now, the probability that any relay can successfully
decode all $N$ packets in the first stage is $P(S) =
\prod_{k=1}^NP\left(|h_{kr}|^2>\tau\right)
=\prod_{k=1}^N\left(1-F(\tau)\right)=\left(1-F(\tau)\right)^N $.
Similar to the definition of $F_m$, let $F_k$ denote the event
that $k$ out of $K$ relays fail to receive all the packets:
$P(F_k) = \binom  Kk P(S)^{K-k}(1-P(S))^k$. Using similar
techniques as in the proof of \thref{th.beta}, it can be shown
that $Pr(E_k)\cong K_1 \tau^{Mk}$ where $K_1=\binom Kk
(2N)^{(M-1)k}\beta^{Mk}N^k $.

Also let $E_{s,t}(k)$ denote the event that $s$ channels out of
$N$ source channels and $t$ channels out of $K-k$ relay channels
were \emph{operational}. Then we have $ P(E_{s,t}(k)) = \binom Ns
P_0^{N-s} (1-P_0)^s \binom {K-k}t F(\tau)^{K-k-t}(1-F(\tau))^t$.
%\begin{align}\label{E_{n}(k)}
% P(E_{s,t}(k)) = \binom Ns  P_0^{N-s}
%            (1-P_0)^s \binom {K-k}t F(\tau)^{K-k-t}(1-F(\tau))^t
% \end{align}
Notice that, since $\Gamma_N(A)=N$ we have $E_i=E_i^{up}$.
Therefore, we have $P(E_{i}) = \sum_{k=0}^{K} {P(F_k)} \cdot
  {\sum_{\{s,t|s+t=0\}}^{N-1} P(E_{s,t}(k))}$.
%\begin{align}
%  P(E_{i}) = \sum_{k=0}^{K} {P(E_k)} \cdot
%  {\sum_{\{s,t|s+t=0\}}^{N-1} P(E_{s,t}(k))}
% \end{align}
It can be shown that $P(E_{s,t}(k)) \cong K_2
\tau^{(N-s)^++(K-k-t)^+M}$, where $(x)^+=\max(x,0)$ and
$K_2=\binom Ns \binom {K-k}t
\beta^{N-s+(K-k-t)M}(2N)^{(M-1)(K-k-t)}$
%  \begin{align}
%   \lim_{\tau \rightarrow 0} {\frac{P(E_{s,t}(k))}{\tau^{N-s+(K-k-t)M}}} =
%    \binom Ns \binom  {K-k}t \beta^{N-s+(K-k-t)M}(2N)^{(M-1)(K-k-t)}
%  \end{align}
and hence $P(E_{i}) \cong K_1K_2\tau^{Mk+(N-s)^++(K-k-t)^+M}$.
%  \begin{align}
%   \lim_{\tau \rightarrow 0} {\frac{P(E_{i})}{\tau^{Mk+N-s+(K-k-t)M}}} =
%& \binom Kk (2N)^{(M-1)k}\beta^{Mk}N^k \cdot \nonumber\\
% & \binom Ns \binom{K-k}t \beta^{N-s+(K-k-t)M}(2N)^{(M-1)(K-k-t)}
%  \end{align}
We need to find out $\min_{\{s,t|s+t=\{0, \ldots,
N-1\}\}}{((N-s)^++(K-t)^+M)}$.
%\begin{equation}
%\min_{\{s,t|s+t=\{0, \ldots, N-1\}\}}{((N-s)^++(K-t)^+M)}
%\end{equation}
We need to consider two different cases: $K<N-1$ and $K\geq N-1$.
For the first case, choosing $t=K$ and $s=N-1-t=N-1-K$ achieves
the minimum: $(N-s)^++(K-t)^+M=(N-(N-1-K)+(K-K)M=K+1$. And for the
second case choosing $t=N-1$ and $s=0$ achieves the minimum:
$(N-s)^++(K-t)^+M=N+(K-(N-1))M$. Now, rest of the proof can be
completed using similar techniques as in the proof of
\thref{th.beta}.

\end{proof}

In \cite{bkrl06}, for a single source single destination setup it
was proved that instead of transmitting from all the $M$ relays,
if a selection is performed and only the relay with the best
channel coefficient transmits then the BER at the destination
enjoys a $M$-fold diversity gain. Inspired by this idea, the
authors in \cite{pzzy08} proposed a Network-coded cooperation
(NCC) scheme for $N$ s-d pairs and $M$ relays where only the
``\emph{best}'' relay selected according to \eqref{selectionrule},
XOR'es all the source packets and transmits to the destination
(\figref{fig.ta} (b)). However, the $M$-fold diversity order can
only be achieved when an unrealistic assumption is made. The
assumption is that the destination has to be able to decode all
the other source packets successfully. When no such assumption is
made, no gain from the selection process is obtained and only a
diversity order of one is achieved. The significance of our result
in \thref{extension} is that if enough number of relays could be
used ($K \geq N-1$), we can achieve $N+M\left(K-(N-1)\right)$
diversity order.

\section{Comparison with Other Schemes} \label{sec.compare}
\subsection{DMT Comparison}

In this section, we would like to compare diversity-multiplexing
tradeoff of the previously proposed schemes in the literature. The
closest scheme in the literature is the NCC scheme considered in
\cite{pzzy08}. In NCC instead of all the relays, only one relay
transmits which results in total of $N+1$ time slots. Using fewer
time-slots NCC achieves a better spectral efficiency than the
proposed scheme here. However, NCC can only provide a fixed
diversity order of two, while the the proposed scheme achieves the
full-diversity order of $M+1$.

In the following, for comparison we include the DMT performance of
NCC and that of conventional cooperation (CC) which includes
space-time coded protocols \cite{lawo03} and selection relaying
\cite{bkrl06}. The diversity-multiplexing tradeoff of NCC is given
by \cite{pzzy08}: $d\left(r\right) = 2\left(1-r(N+1)/N\right) , \;
r \in \left(0,\frac{N}{N+1}\right)$. The DMT of the decode and
forward strategy with $M$ intermediate relay nodes is given by
\cite{lawo03}: $d\left(r\right) =
\left(M+1\right)\left(1-2r\right) , \; r \in \left(0,0.5\right).$

\begin{figure}
\centering
\includegraphics[width=.65\textwidth]{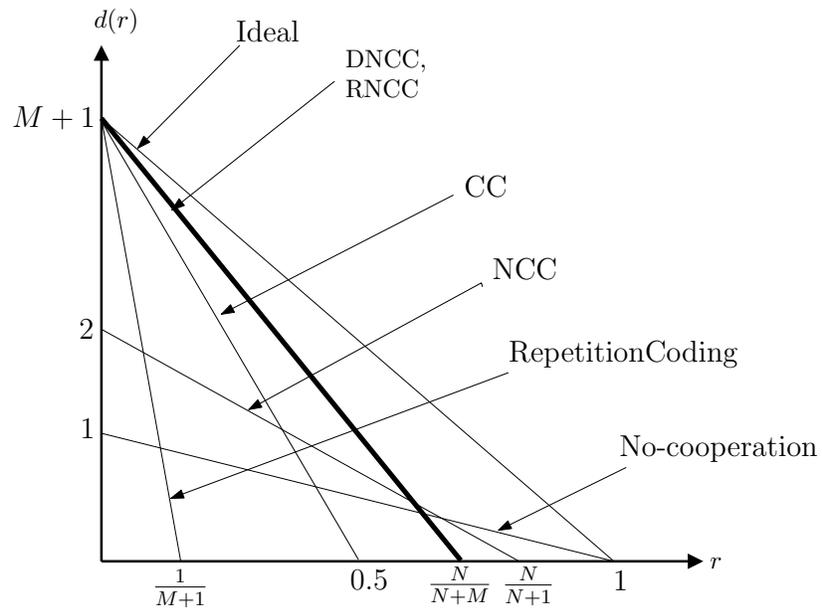}
\caption{DMT comparison of various schemes} \label{fig.dmt}
\end{figure}

To show the advantage of the proposed schemes, we present
diversity-multiplexing tradeoff of the existing schemes and the
proposed schemes in \figref{fig.dmt}. As can be seen from the
figure, both of the proposed schemes and CC provide a full
diversity order of $M+1$ when $r\rightarrow 0$. However, the
proposed schemes can provide a higher diversity gain than CC when
the spectral efficiency increases.

\subsection{System Outage Probability}

\begin{figure}
\centering
\includegraphics[width=.75\textwidth]{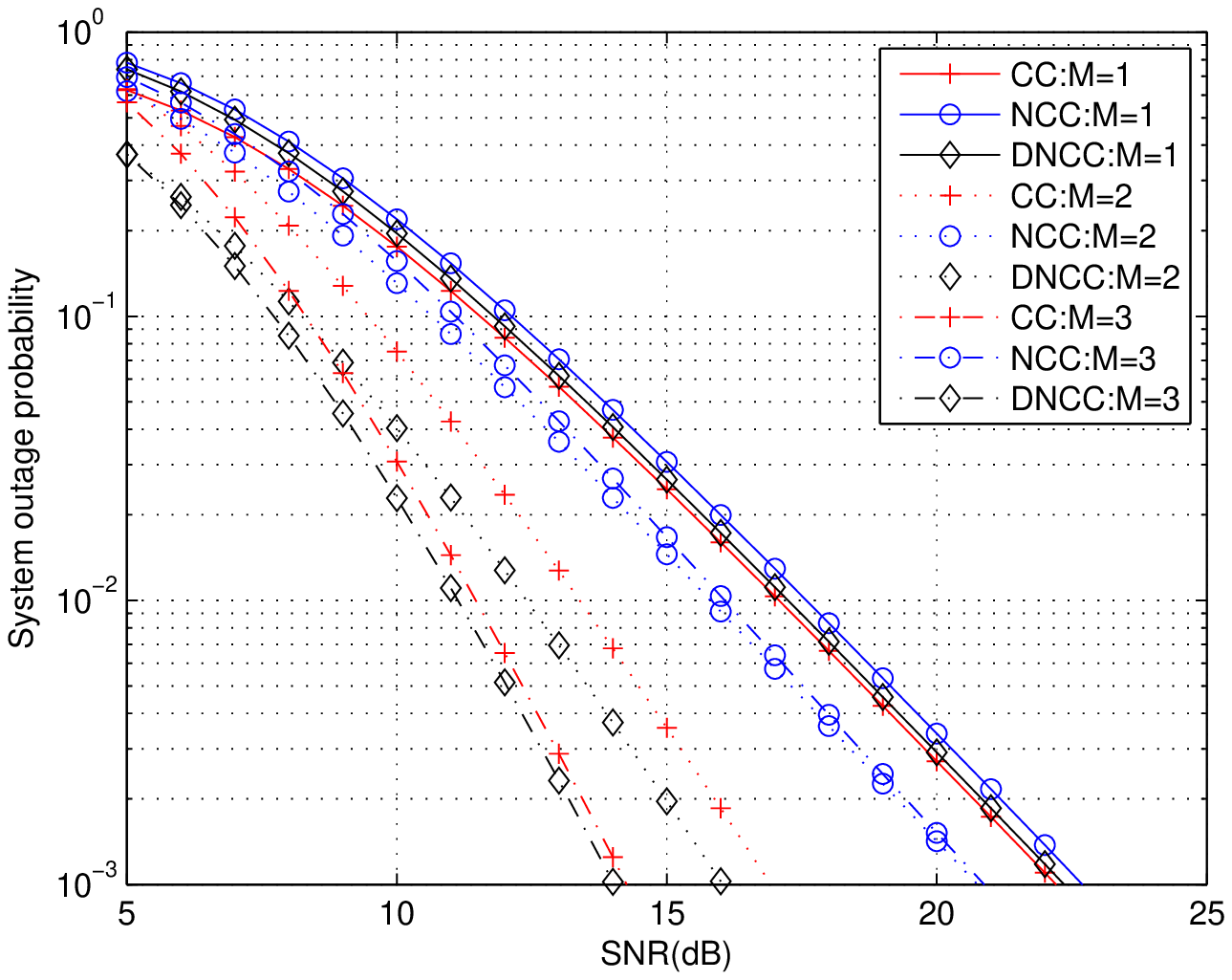}
\caption{System outage probability comparison. (N=2 M=1, \ldots
3)} \label{fig.outN2R1}
\end{figure}

\begin{figure}
\centering
\includegraphics[width=.75\textwidth]{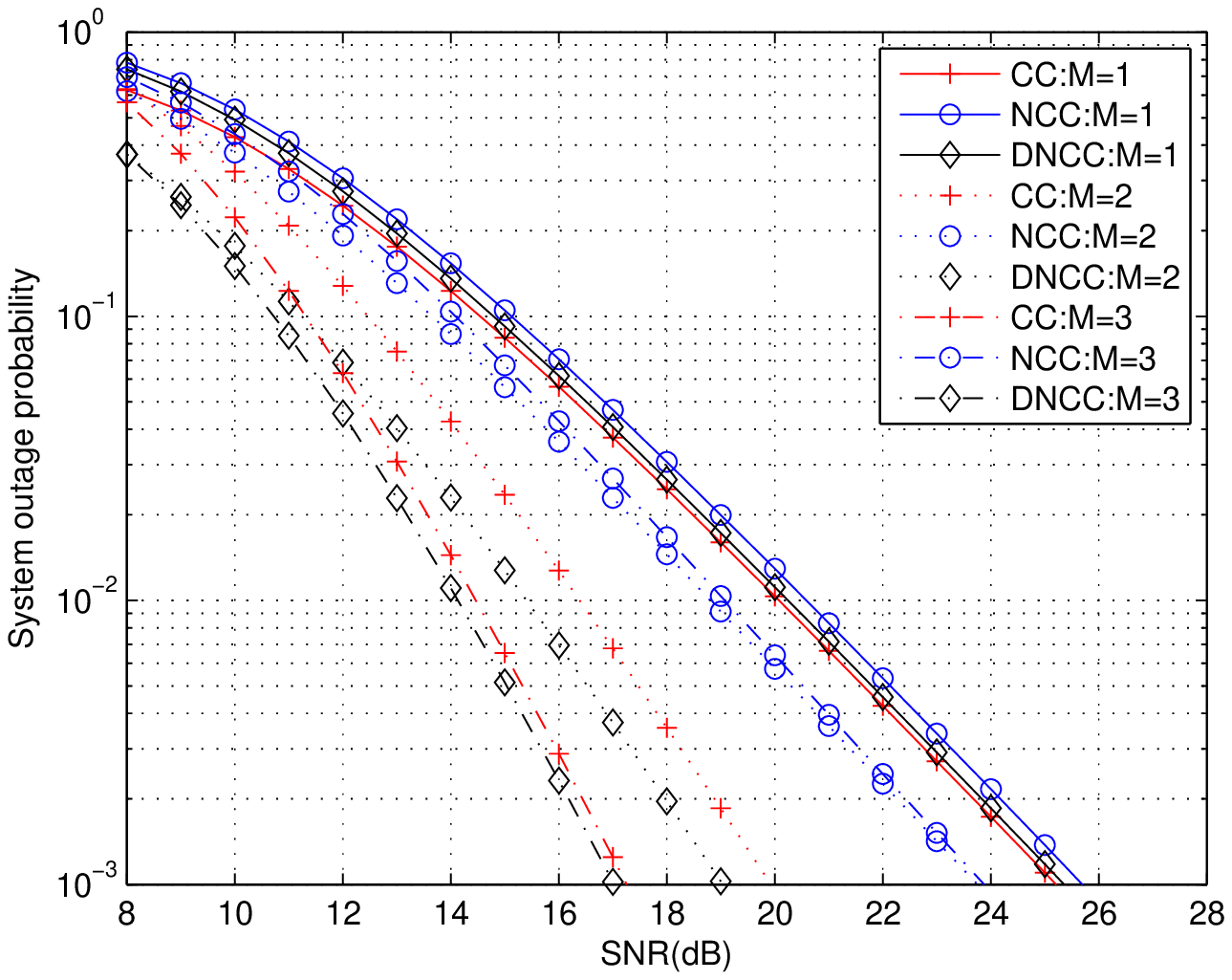}
\caption{System outage probability comparison. (N=3, M=1, \ldots
3)} \label{fig.outN3R1}
\end{figure}

Here, we compare the system outage probability of the proposed
schemes with the other schemes. The system outage occurs when
\emph{any} $d_i$ is unable to decode $\Theta_i$ reliably: $P_s =
1-\prod_{i=1}^N{(1-P(E_{i}))}$.
%\begin{equation}\label{sysout}
%P_s = 1-\prod_{i=1}^N{(1-P(E_{i}))}
%\end{equation}
We compare $P_s$ with the system outage probabilities (30), (43)
derived in \cite{pzzy08}. In \figref{fig.outN2R1} for a network
consisting of two source-destination pairs with various number of
relays is compared. As can be seen from the figure, the proposed
method clearly outperforms NCC by achieving the full diversity of
$M+1$ as compared to NCC's fixed diversity order of 2. The same
performance is observed in \figref{fig.outN3R1} where the number
of source-destination pairs is increased to three.

\subsection{Monte-Carlo Simulation}
\begin{figure}
\centering
\includegraphics[width=.75\textwidth]{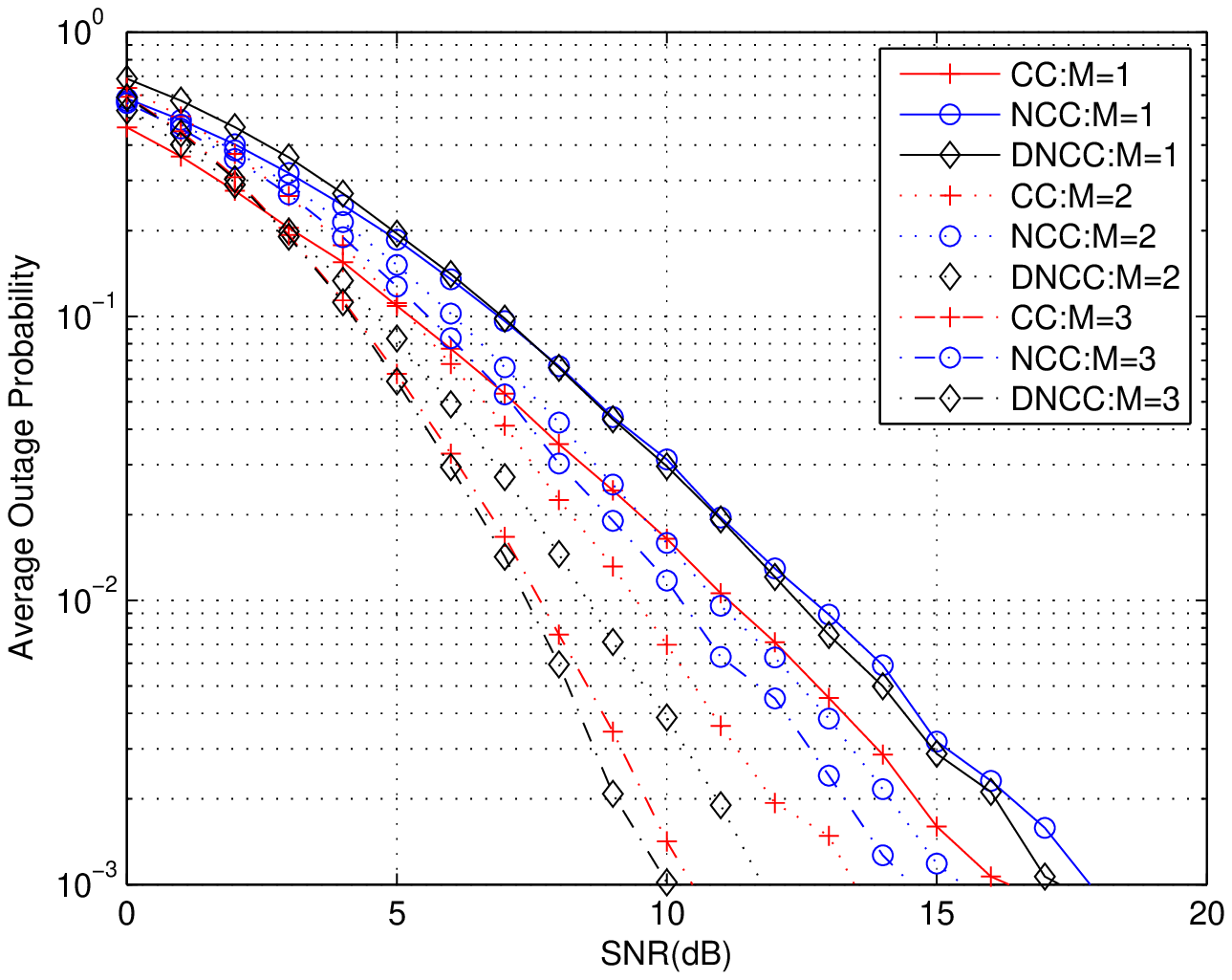}
\caption{Average outage probability per destination, (N=2, M=1,
\ldots 3)} \label{fig.monteN2R1}
\end{figure}

\begin{figure}
\centering
\includegraphics[width=.75\textwidth]{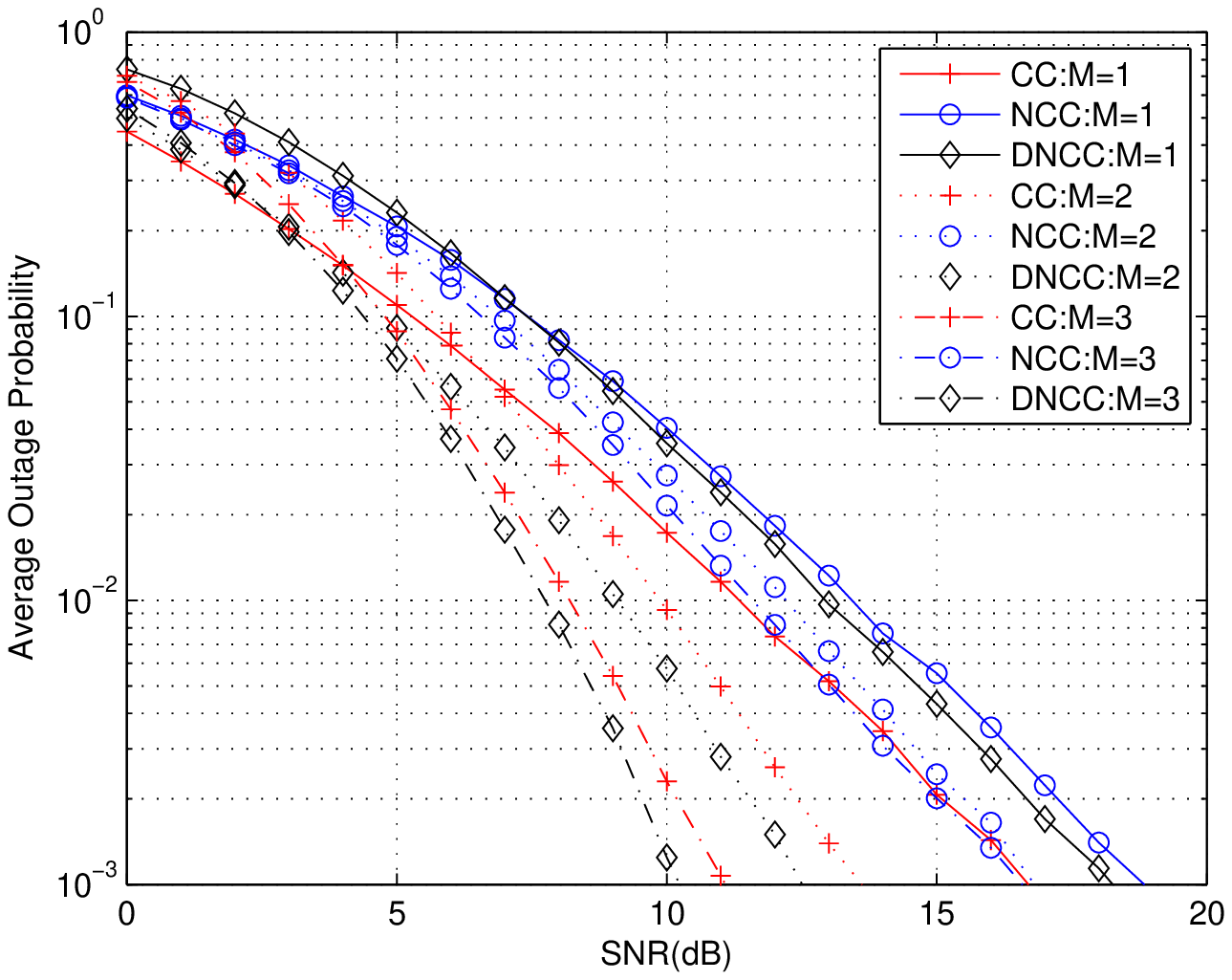}
\caption{Average outage probability per destination, (N=3, M=1,
\ldots 3)} \label{fig.monteN3R1}
\end{figure}

Here, we compare these schemes with the existing schemes via
Monte-Carlo simulations for various number of relays. In the
simulations, only channel conditions are considered to isolate the
diversity benefits of the scheme. We generate an $(N+M) \times N$
and an $N \times N$ matrix that contains the channel coefficients
for each destination and each relay, respectively. Then, we decide
that the transmission is successful for any link if the
instantaneous channel condition is large enough to be able to
support the given data rate and we update the same size linear
coefficient matrices accordingly. After all the transmissions take
place, we perform Gaussian elimination on the updated linear
coefficient matrices to conclude whether each destination $d_i$
was able to recover the source packet $\Theta_i$ or not. The
channel coefficient variances are chosen to be equal to one and
$R_0=1$ BPCU. Please note that we considered the average outage
error probability which is found by dividing the total number of
errors occurred by the number of source nodes instead of the
system error probability. Please note that, in all the figures
only the unicast scenario is adapted since CC cannot be
implemented in a multicast scenario. We compare the proposed
schemes in \figref{fig.monteN2R1} and \figref{fig.monteN3R1} with
NCC and CC. As can be seen from the figures, the proposed schemes
are able to provide the $M+1$ diversity order and outperform the
other schemes.

\begin{figure}
\centering
\includegraphics[width=.75\textwidth]{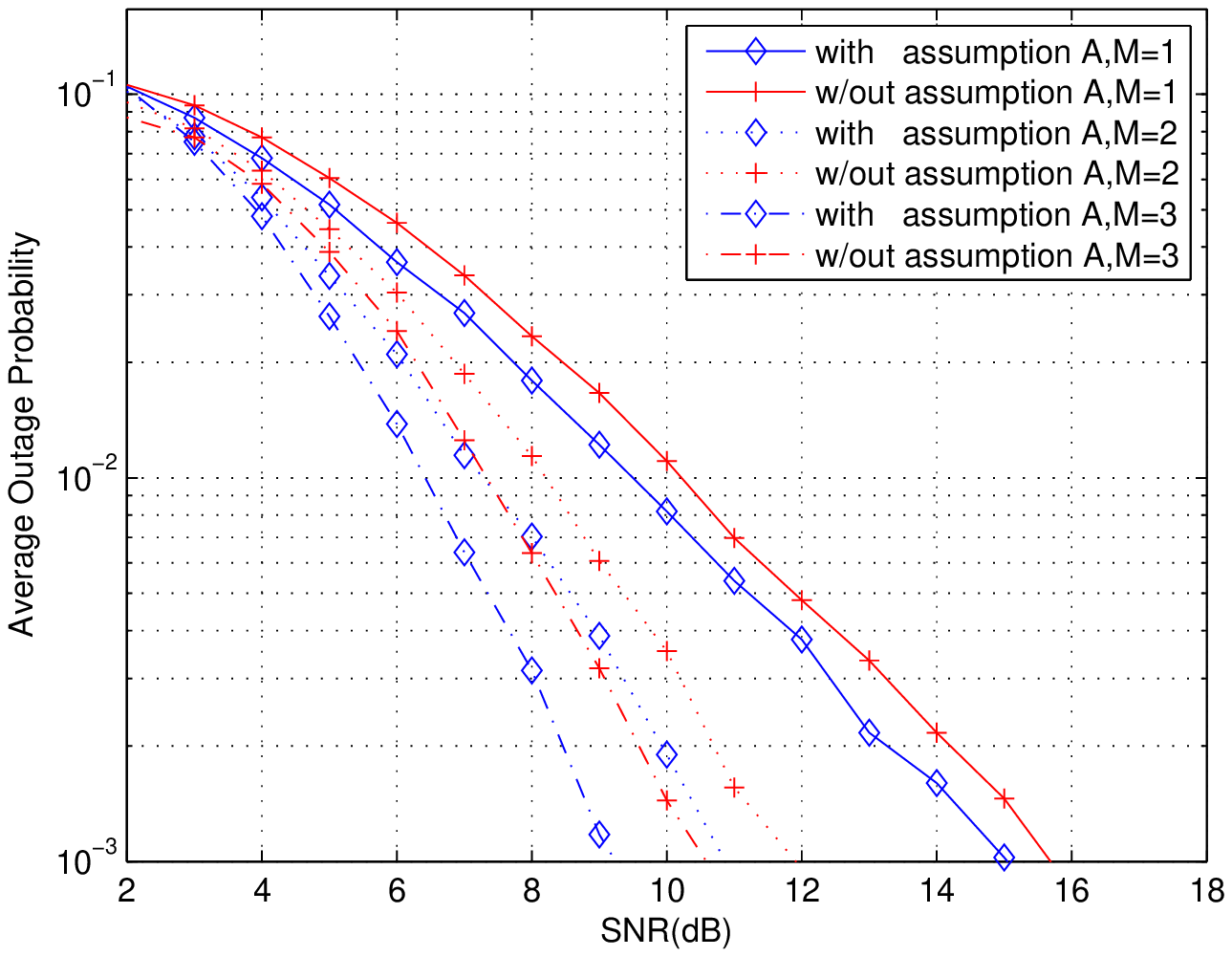}
\caption{Average outage probability per destination, (N=2 M=1,
\ldots 3)} \label{fig.decodeallN2}
\end{figure}

\begin{figure}
\centering
\includegraphics[width=.75\textwidth]{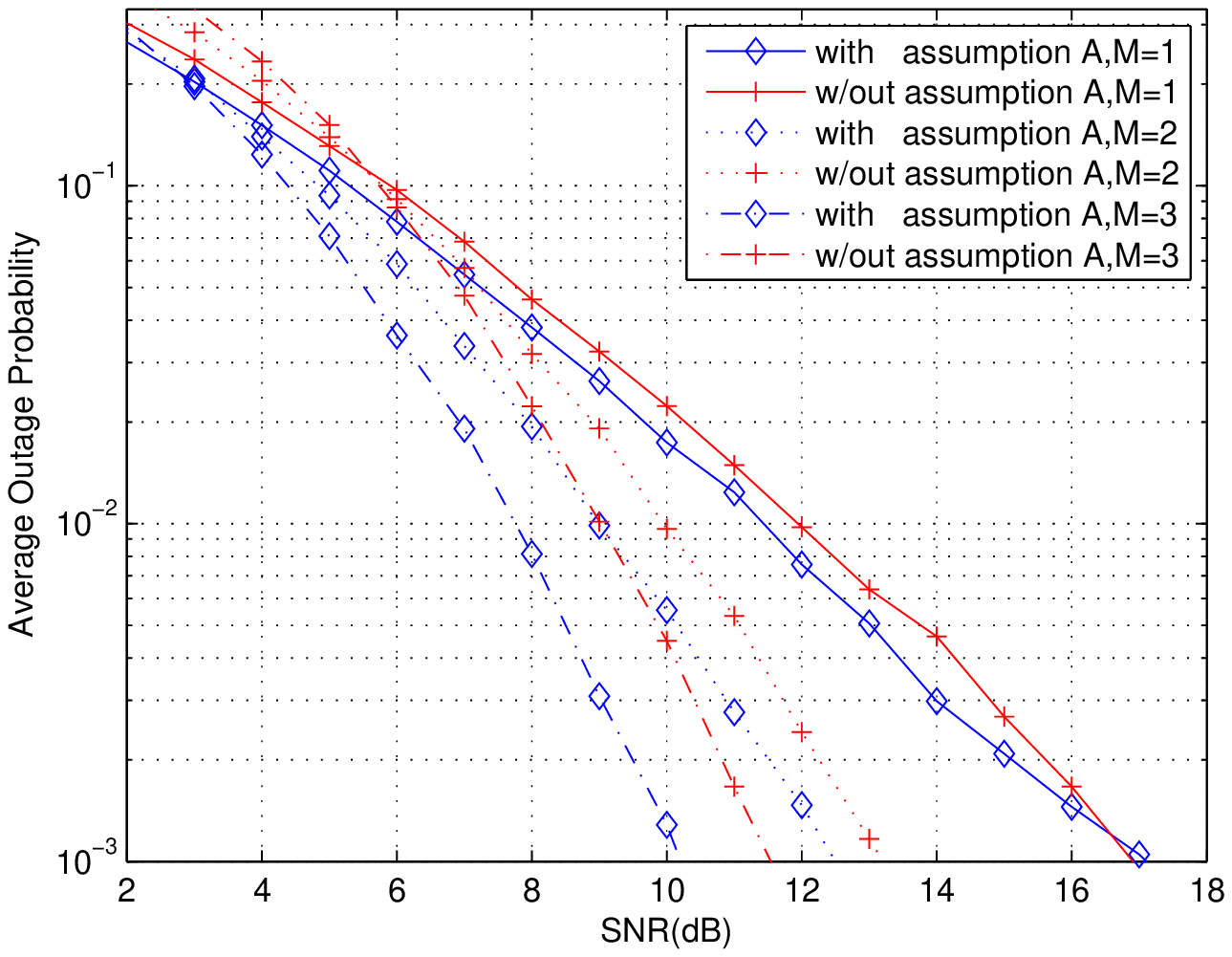}
\caption{Average outage probability per destination, (N=3 M=1,
\ldots 3)} \label{fig.decodeallN3}
\end{figure}

%We also validate the results of \thref{th.maindmt} through
%simulations.
\thref{th.maindmt} claims that the performance loss incurred due
to the assumption under Strategy $\cal{A}$ is not in terms of
diversity gain but it is in terms of coding gain. This is
validated through simulations as shown in \figref{fig.decodeallN2}
and \figref{fig.decodeallN3} using Monte-Carlo simulations which
are set up in the same way as in \figref{fig.monteN2R1} and
\figref{fig.monteN3R1}.

\section{Conclusions} \label{sec.conclusions}

In this paper, we have proposed two network coded cooperation
schemes for $N$ source-destination pairs assisted with $M$ relays.
We studied two different traffic network models: multicast and
unicast. The proposed schemes allow the relays to apply network
coding on the data it has received from its neighbors. We allow
the relays to linearly combine the packets with coefficients
either drawn randomly from a finite field or drawn from a linearly
independent set of coefficients. We established the
diversity-multiplexing tradeoff performance of the proposed
schemes, and showed its advantage over the existing schemes.
Specifically, it achieves the full-diversity order $M+1$ at the
expense of a slightly reduced multiplexing rate. The novelty in
the proposed scheme is that we establish the link between the
parity-check matrix for a $(N+M,M,N+1)$ systematic MDS code and
the coefficients to perform network coding in a cooperative
communication scenario consisting of $N$ source-destination pairs
and $M$ relays. We presented two ways to generate the network
coding matrix: using the Cauchy matrices and the Vandermonde
matrices.

%\begin{appendices}
%\end{appendices}
\bibliographystyle{IEEE}
\bibliography{refs}

\newpage

\newpage

\newpage

\newpage

\newpage

\newpage

\newpage

\newpage

\newpage

\end{document}